\documentclass[twocolumn,showpacs,prl]{revtex4}

\usepackage{graphicx}
\usepackage{dcolumn}
\usepackage{bm}

\begin{document}

\preprint{???}

\title{Collective Dynamics of Lipid Membranes studied by Inelastic Neutron Scattering}

\author{M.C.~Rheinst\"adter$^{1,2}$ \email{rheinstaedter@ill.fr},
C.~Ollinger$^3$, F.~Demmel$^1$, G.~Fragneto$^1$, and T.~Salditt$^3$}

\affiliation{$^1$Institut Laue-Langevin, 6 rue Jules Horowitz, BP
156,38042 Grenoble Cedex 9, France \\
$^2$Institut f\"{u}r Festk\"{o}rperforschung,
Forschungszentrum-J\"{u}lich, 52425
J\"{u}lich, Germany \\
$^3$Institut f\"{u}r R\"{o}ntgenphysik,
Georg-August-Universit\"{a}t G\"{o}ttingen, Geiststra{\ss}e 11,
37037 G\"{o}ttingen, Germany }

\date{\today}

\begin{abstract}
We have studied the collective short wavelength dynamics in
deuterated DMPC bilayers by inelastic neutron scattering. The
corresponding dispersion relation $\hbar\omega$(Q) is presented
for the gel and fluid phase of this model system. The temperature
dependence of the inelastic excitations indicates a phase
coexistence between the two phases over a broad range and leads to
a different assignment of excitations than that reported in a
preceding inelastic x-ray scattering study [Phys. Rev. Lett. {\bf
86}, 740 (2001)]. As a consequence, we find that the minimum in
the dispersion relation is actually deeper in the gel than in the
fluid phase. Finally, we can clearly identify an additional
non-dispersive (optical) mode predicted by Molecular Dynamics (MD)
simulations [Phys. Rev. Lett. {\bf 87}, 238101 (2001)].
\end{abstract}


\maketitle

The collective dynamics of lipid molecules are believed to affect
significantly the physical properties of phospholipid membranes
\cite{Lipowsky:1995,Bayerl:2000}. In the context of more complex
biological membranes, collective molecular motions may play a
significant role for different biological functions. For example,
correlated molecular motions of the lipid acyl chains and the
corresponding density fluctuations in the plane of the bilayer are
believed to play an important role for the transport of small
molecules through the bilayer \cite{Paula:1996}. Molecular
vibrations, conformational dynamics and 'one particle' diffusion
in the plane of the bilayer can be studied by a number of
different  spectroscopic techniques covering a range of different
time scales such as incoherent inelastic neutron scattering
\cite{Pfeiffer:1989} or nuclear magnetic resonance
\cite{Nevzorov:1997}. Contrarily, few experimental techniques are
able to elucidate the short range collective motions mentioned
above. To this end, Chen {\em et al.\@} have recently presented a
seminal inelastic x-ray scattering (IXS) study
\cite{Chen:2001,Weiss:2003} of the dispersion relation $\hbar
\omega$(Q$_r$) which quantifies the collective motion of the lipid
acyl chains as a function of the lateral momentum transfer Q$_r$.
The basic scenario is the following: At small Q$_r$, longitudinal
sound waves in the plane of the bilayer are probed and give rise
to a linear increase  of $\omega\propto$ Q$_r$, saturating at some
maximum value ('maxon'), before a pronounced  minimum $\Omega_0$
('roton') is observed at Q$_0\simeq$ 1.4 \AA$^{-1}$, the first
maximum in the static structure factor S(Q$_r$) (the inter-chain
correlation peak). Qualitatively, this can be understood if Q$_0$
is interpreted as the quasi-Brillouin zone of a two-dimensional
liquid. Collective modes with a wavelength of the average nearest
neighbor distance $2\pi$/Q$_0$ are energetically favorable,
leading to the minimum. At Q$_r$ values well above the minimum,
the dispersion relation is dominated by single particle behavior.
A quantitative theory which predicts the absolute energy values of
'maxon' and 'roton' on the basis of molecular parameters is absent
so far. However, the dispersion relation can be extracted  from
Molecular Dynamics (MD) simulations by temporal and spatial
Fourier transforming the molecular real space coordinates
\cite{Tarek:2001}.

\begin{figure}
\centering
\resizebox{1.00\columnwidth}{!}{\rotatebox{0}{\includegraphics{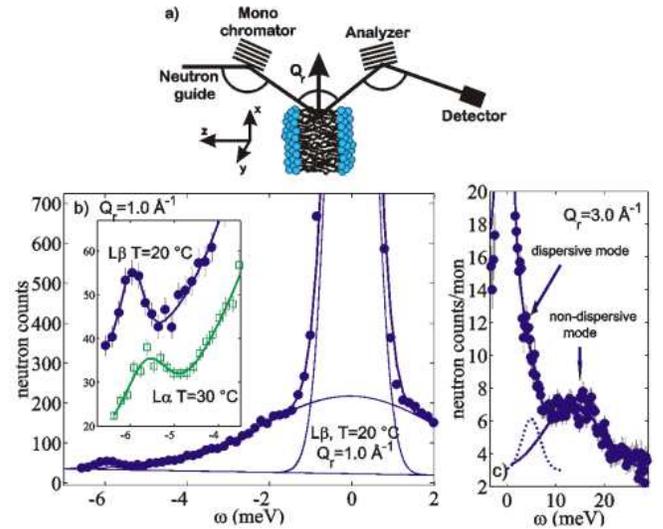}}}
\caption[]{(a) Schematic of the scattering geometry. (b) Energy
scan in the gel phase of DMPC (T=20 $^{\circ}$C), measured at
Q$_r$=1.0 \AA$^{-1}$ (IN12 data). The scattering is composed of
the central elastic peak, the broad quasi-elastic background, and
symmetric satellites indicating the excitation. The inset shows a
zoom of the satellite peaks in both gel and fluid phase (T=30
$^{\circ}$C). (c) Scan over a higher energy transfers range
(Q$_r$=3.0 \AA$^{-1}$, IN3 data), showing both the dispersive
excitation and the non-dispersive (optical) excitation.}
\label{q100_final.eps}
\end{figure}

Here we present a first time inelastic neutron scattering (INS)
experiment on the collective dynamics of the lipid acyl chains in
the model system DMPC -d54. Note that by selective deuteration of
the chains, the respective motions are strongly enhanced over
other contributions to the inelastic scattering cross section. We
have measured the dynamical structure factor S(Q$_r,\omega)$ in
the gel (L$_{\beta}$) and fluid (L$_{\alpha}$) phases, and have
investigated the temperature dependence of the excitations in the
dispersion minimum in the vicinity of the main phase transition
\footnote{Note that in the deuterated compound there is no
P$_{\beta'}$ (ripple) phase.}. While qualitatively we obtain
similar curves as Chen {\em et al.\@} \cite{Chen:2001}, several
results are significantly different and in striking contrast to
the earlier study. Notably, the minimum in the dispersion relation
measured by INS is deeper in the gel than in the fluid phase
$\Omega_{0,L\beta} \le \Omega_{0,L\alpha}$, while the opposite is
reported in \cite{Chen:2001}. Secondly, the INS data give much
smaller peak widths of around 1 meV, indicating a much smaller
damping, while Chen {\em et al.\@} report values in the range of
$\Delta\omega_s=4-6$ meV. Finally, we observe a second
non-dispersive mode at higher energy transfer of 14 meV, in
addition to the dispersive sound mode. This mode was not observed
by Chen {\em et al.\@}, possibly due to the low signal-to-noise
ratio in IXS at high energy transfer. It has been predicted by a
recent Molecular Dynamics (MD) simulation  and can be attributed
to the terminal (methyl) carbons of the chains \cite{Tarek:2001}.
The measurements were carried out on the cold triple-axis
spectrometer IN12 and the thermal spectrometer IN3 at the high
flux reactor of the Institut Laue-Langevin (ILL) in Grenoble,
France. The main differences with respect to IXS are related to
the energy-momentum relation of the neutron versus the photon
probe, strongly affecting energy resolution, accessible
(Q,$\omega$) range \footnote{Due to  the dispersion relation of
the neutron itself ($\sim Q^2$), and contrarily to IXS, the range
at low $Q$ and high $\omega$ values is not accessible by INS.} and
the signal-to-noise ratio. In the present case, the energy of the
incident neutrons was in the range of the excitations (some meV)
resulting in a high energy resolution up to $\sim$300 $\mu$eV, in
comparison to 1.5 meV of the IXS experiment.  A better energy
resolution in combination with a smaller ratio between central
peak and Brillouin amplitudes leads to very pronounced satellites,
see Fig. \ref{q100_final.eps}. This is of particular advantage for
the identification of peaks in the central part of the dispersion
relation, as well as for the experimental verification of the
predicted non-dispersive mode at high energies.

Deuterated DMPC -d54 (deuterated
1,2-dimyristoyl-sn-glycero-3-phoshatidylcholine) was obtained from
Avanti Polar Lipids.
Highly oriented membrane stacks were prepared by spreading a
solution of typically 35 mg/ml lipid in TFE/chloroform (1:1) on
2'' silicon wafers, followed by subsequent drying in vacuum and
hydration from D$_2$O vapor \cite{Muenster:1999}.  Nine such
wafers separated by small air gaps were combined and aligned with
respect to each other to create a 'sandwich sample' consisting of
several thousands of highly oriented lipid bilayers (total
mosaicity of about 0.6 deg), with a total mass of about 300 mg of
deuterated DMPC. The samples were kept in a closed temperature and
humidity controlled Aluminum chamber. Hydration of the lipid
membranes was achieved by separately adjusting two heating baths
(Haake, Germany), connected to the sample chamber and to a heavy
water reservoir, hydrating the sample from the vapor phase.
Temperature and humidity sensors were installed close to the
sample. Before the measurements, the samples were equilibrated at
T=55 $^{\circ}$C for about two hours to anneal defects. The
scattering vector, {\bf Q}, was placed in the plane of the
membranes (see Fig. \ref{q100_final.eps} (a) for a sketch of the
scattering geometry) to measure the static structure factor
S(Q$_r$) in the plane of the membranes as well as the dynamic
structure factor S(Q$_r,\omega)$ in the same run without changing
the setup.

A typical energy scan (raw data) is shown in Fig.
\ref{q100_final.eps} (b), taken at T=20 $^{\circ}$C, in the gel
phase of the bilayer at Q$_r$=1.0 \AA$^{-1}$ (all data are
corrected for temperature by the Bose factor). The inset shows the
excitations of the bilayer in the gel and the fluid phase (at T=30
$^{\circ}$C, nine degrees above the phase transition temperature
\footnote{The phase transition from L$_{\beta}$ to L$_{\alpha}$
occurs at T$_c$=21.5 $^{\circ}$C for DMPC -d54.} ). Position and
width can easily be determined from these well pronounced peaks.
The inelastic scans can be evaluated by the generalized three
effective eigenmode theory (GTEE)
\cite{Liao:2000,Chen:2001,Weiss:2003}, using the following
function for least-square fitting
\begin{eqnarray*}
\frac{S(Q,\omega)}{S(Q)}&=&\frac{1}{\pi}\left(A_0\frac{\Gamma_h}{\omega^2+\Gamma_h^2}\right.\\\nonumber
&+&A_s\left[\frac{\gamma_s+b(\omega+\omega_s)}{(\omega+\omega_s)^2+\gamma_s^2}+\left.\frac{\gamma_s-b(\omega-\omega_s)}{(\omega-\omega_s)^2+\gamma_s^2}\right]\right)
\nonumber.
\end{eqnarray*}
The model consists of a heat mode, centered at $\omega=0$ meV
(Lorentzian with a width $\Gamma_h$), and two sound modes,
represented by Lorentzians  at $\omega=\pm\omega_s$ and damping
$\gamma_s$ \cite{Liao:2000,Weiss:2003}. 
From width of the central mode, and width and position of the
Brillouin lines, the thermal diffusivity, the sound frequency, and
the sound damping and can be determined, respectively, within the
framework of a hydrodynamic theory. To fit the neutron data, we
have to add an additional Lorentzian component describing the
broad quasielastic contribution associated with intramolecular
degrees of freedom and incoherent scattering, not seen by IXS. The
solid line in Fig. \ref{q100_final.eps} (b) is a fit to the GTEE
model with an additional Lorentzian component. Figure
\ref{q100_final.eps} (c) shows an energy scan in the gel phase at
Q$_r$=3.0 \AA$^{-1}$, up to an energy transfer of 30 meV. Aside
from the dispersive excitation due to in-plane density waves, a
second non-dispersive (optical) mode is observed at about
$\omega$=14 meV with a width (FWHM) of about 13 meV, corresponding
exactly to the predictions by Tarek {\em et al.\@}
\cite{Tarek:2001}, and attributed to the methyl ends of the acyl
chains.

\begin{figure}
\centering
\resizebox{1.00\columnwidth}{!}{\rotatebox{0}{\includegraphics{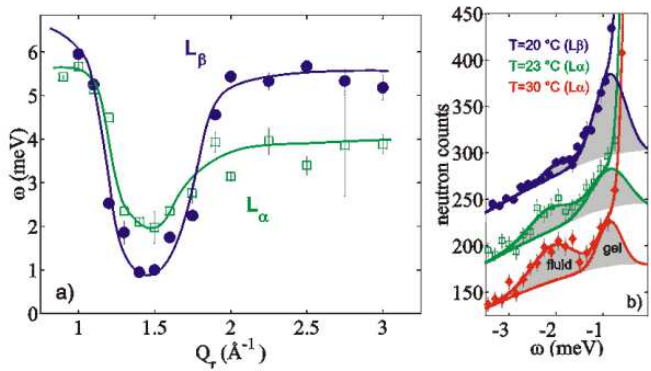}}}
\caption[]{(a) Dispersion relations in the gel and the fluid phase
of the DMPC bilayer. (b)  Energy scans in the dispersion minimum
at Q=1.5 \AA$^{-1}$ for temperatures T=20, 23 and 30 $^{\circ}$C
.}\label{dispersion}
\end{figure}
The dispersion relation in the gel and the fluid phase is shown in
Fig. \ref{dispersion} (a), after evaluating the energy positions
of the satellite (Brillouin) peaks measured in a series of
constant Q-scans in the range from Q=0.7 \AA$^{-1}$ to Q=3.0
\AA$^{-1}$. The fluid dispersion has been measured at T=30
$^{\circ}$C, far above the phase transition and the regime of
so-called anomalous swelling in the fluid phase of the DMPC
bilayers. Fig. \ref{dispersion} (b) shows energy scans in the
dispersion minimum, measured with an enhanced energy resolution.
These curves cannot be fitted by only one excitation. Rather two
contributions have to be assumed, which we identify from their
respective T-dependence, as discussed below. As a consequence, the
dispersion relation in the minimum goes down to smaller values in
the gel than in the fluid phase, distinctly different from
\cite{Chen:2001}. While the 'maxon' and the high-Q range are
energetically higher in the gel than in the fluid phase (due to
stiffer coupling between the lipid chains in all-trans
configuration), $\Omega_0$, the energy value in the dispersion
minimum, is actually smaller in the gel phase, roughly analogous
to soft modes in crystals.

\begin{figure}
\centering
\resizebox{1.00\columnwidth}{!}{\rotatebox{0}{\includegraphics{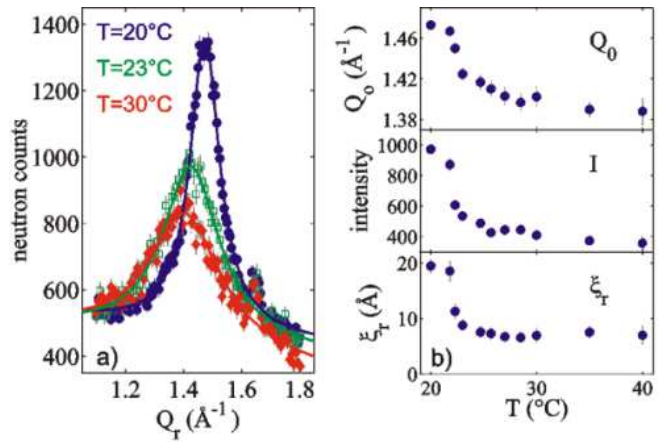}}}
\caption[]{(a): Elastic structure factor showing the inter-chain
correlation peak, measured at temperatures T=20, 23 and 30
$^{\circ}$C. (b): Peak position, intensity and correlation length
as fitted for all measured temperatures.}\label{kettenpeak}
\end{figure}

To investigate the temperature dependence of the collective
dynamics, we  carried out simultaneously inelastic and elastic
measurements in the most interesting range around the minimum in
the dispersion relation and correspondingly the maximum of the
(elastic) chain correlation. Temperature was varied between T=20
and 40 $^{\circ}$C, to cover the range of anomalous
pseudo-critical behavior. Although the phase transition is of
first order \cite{Chen:1997,Nagle:1998}, significant (continuous)
changes (anomalies) are observed close to the phase transition in
many thermodynamic quantities, e.g. in the equilibrium distance
$d$ (so-called anomalous swelling). The reason for this phenomenon
is still a matter of debate. To illustrate the characteristic
changes in the (static) chain correlation peak, elastic scans at
temperatures of T=20 , 23 and 30 $^{\circ}$C are plotted in Fig.
\ref{kettenpeak} (a). Upon heating from the gel to the fluid
phase, the peak position shifts to smaller $Q_r$-values (larger
average next neighbor distances) and the peak broadens, indicating
a less well ordered packing of the acyl chains \cite{Spaar:2003},
which is quantified in  Fig. \ref{kettenpeak} (b) in terms of
fitted peak position Q$_0$(T), peak intensity and  correlation
length $\xi_{r}$(T)=1/HWHM. 

Over the same T-range, Fig. \ref{dispersion} (b) shows
representative energy scans taken at constant Q$_r=1.5$
\AA$^{-1}$, in the dispersion minimum. In the gel phase at T=20
$^{\circ}$C we find an excitation at $\omega_s$=-1 meV, which we
associate with the gel phase. At higher temperatures, the gel
excitations decreases in intensity, while a second excitation at
energy values of $\omega_s$=-2 meV is found to increase with T and
can thus be attributed to the fluid phase. As the excitations are
symmetric around the central peak we are sure not to deal with
spurious effects. The assignment of the excitations to the
particular phases is justified by their temperature dependence: In
each phase there is a dominant excitation. Both modes are clearly
dispersive and change energy position when moving out of the
minimum (see Fig. \ref{dispersion} (a)). We find small traces of
the 'fluid excitation' already in the gel phase. At T=30
$^{\circ}$C, far in the fluid phase, there is still the 'gel
excitation', indicating a coexistence of fluid and gel domains.
This coexistence is only observed in the range of the dispersion
minimum which coincides with the maximum of the static structure
factor. Note that at the same time, the elastic scans do not show
coexistence of two phases, but the typical well known behavior:
While the transition is of first order, a pseudo-critical
swelling, i.e.\@ a continuous change of the interlamellar distance
above T$_c$, is observed for DMPC and other lipids
\cite{Nagle:1998,Chen:1997,Mason:2001}. The changes in $d$ are
accompanied by corresponding changes in the mean distance between
chains, see Fig. \ref{kettenpeak} (b). For both quantities,  the
elastic diffraction data show a continuous change, while the
present interpretation of the inelastic curves indicates a
coexistence of gel and fluid domains, in particular also far above
the transition in the fluid phase. A crucial point is the size of
these domains: The coexistence of macroscopic domains with sizes
larger than the coherence length $\xi_n$ of the neutrons in the
sample would lead to a peak splitting of the chain peak.
Therefore, the domain sizes must be smaller than a few hundred
\AA\ estimated for $\xi_n$. Fluid domains in the gel phase and
vice versa with sizes smaller than 0.01 $\mu$m$^2$ have been
reported in a recent AFM study \cite{Xie:2002}, and have been
related to lateral strain resulting from density differences in
both phases.

In summary, the high resolution neutron measurements show that the
dip in the dispersion relation is actually deeper in the gel than
in the fluid phase, according to the assignment of the peaks in
e.g.\@ Fig.\ref{dispersion} (b) as based on their T-dependence. In
this case, the present data also has important implications for
the phase transition from the gel to fluid phase and respective
phase coexistence in lipid bilayers. Finally, we argue that the
minimum $\Omega_0$ must be smaller in the gel phase, since in the
framework of the GTEE model it can be shown that the frequency of
the excitations is approximately equal to $\Omega = Q_r v_0 /
\sqrt{2 S(Q_r)}$ (valid in the region of the minimum, see
\cite{Boon:1980,Chen:2001}), where $v_0 = \sqrt{k_B T /m_s}$ is
the thermal velocity of the scatterers \footnote{Note that $v_0$
must obviously be related to an effective mass of the scatterers
$m_s$. It could possibly also be a $Q_r$-dependent quantity, even
though no $Q_r$ dependence results from the form factor like in
IXS.}. If this relation holds, it is clear that the minimum is
deeper for the gel phase, where the elastic scattering S$(Q_r)$ is
more strongly peaked. Let us illustrate the relevant length and
time scales of the excitations. With the given values of
$\Omega_0$ and the corresponding wave vector Q$_0$, we can
estimate the wave velocity of the corresponding propagating modes
in the two phases: $c_{L\beta} = \Omega_0 /Q_0 \simeq 100$ m/s and
$c_{L\alpha} \simeq 220$ m/s. From the measured width of the
excitation, we estimate the lifetime
$\tau=2\pi/\Delta\omega_s\simeq$ 17 ps and 9.2 ps, as well as the
dynamic correlation length of the excitations $\xi_{\tau} = c \tau
\simeq 17$ \AA\ and $20$ \AA, for the gel and fluid phase,
respectively. These length scales are almost identical for the two
phases, and fall very close to the static correlation length
$\xi_{r,\beta}\simeq 20$\AA\ in the gel phase.

It can be speculated that the dispersion relation in both phases
is important for the transport of molecules across the bilayer as
well as parallel to the bilayer. In particular the self-diffusion
of the lipid molecules  may be coupled to the collective motions
of the lipid tails. Free area needed for diffusion might be
generated by density waves which are incommensurate with the mean
inter-molecular spacing, $2\pi/Q_0$. The frequency $\Omega
/(2\pi)$ of the collective density  waves may serve as trial
frequencies for thermally activated chain conformations which are
involved in the formation of free volume. The free volume model of
diffusion gives a diffusion constant $D=D'\exp({ - a_*/a_f
-E_a/(k_B T)})$ , which depends on the critical free area $a_*$
per molecule needed for a hopping process, the average free area
per molecule $a_f$ and the energy barrier $E_a$ for diffusion
\cite{Almeida:1992,Almeida:1995}. Possibly, trial jump rates
invoked in the model (in the pre-exponential term) may be related
to the frequency of collective motions. Furthermore, the energy of
the collective chain motions measured here, must be related to the
average amplitude of excitations at thermal equilibrium, these
amplitudes may determine the probability for the generation of
critical free volume. Certainly, more work both experimental and
theoretical is needed to investigate the relationship between
self-diffusion and collective chain motions. To this end, studying
elementary diffusion events by molecular dynamics simulation may
provide important insight.

Acknowledgements: We thank the ILL for allocation of ample beamtime. This
work has been funded by the German Research Ministry under
contract numbers 05300CJB6 (M.C.R.) and 05KS1TSA7 (C.O. and T.S.).

\bibliography{./dmpc_PRL_etal}

\end{document}